%% file: main.tex
\documentclass[journal]{IEEEtran}
\usepackage{color}

\usepackage{cite}

\ifCLASSINFOpdf
\else
\fi
\usepackage{amsmath}
\usepackage{array}

\ifCLASSOPTIONcompsoc
 \usepackage[caption=false,font=normalsize,labelfont=sf,textfont=sf]{subfig}
\else
 \usepackage[caption=false,font=footnotesize]{subfig}
\fi

\ifCLASSOPTIONcaptionsoff
 \usepackage[nomarkers]{endfloat}
\let\MYoriglatexcaption\caption
\renewcommand{\caption}[2][\relax]{\MYoriglatexcaption[#2]{#2}}
\fi
\usepackage{url}
\usepackage{hyperref}
\usepackage{mathtools}
\usepackage{epstopdf}
\usepackage{multirow}
\usepackage[utf8]{inputenc}

\begin{document}
%

\title{Fleet-Level Environmental Assessments for Feasibility of Aviation Emission Reduction Goals}

%

\author{Kolawole~Ogunsina,
		Hsun~Chao,
        Nithin~Jojo~Kolencherry,
        Samarth~Jain,
        Kushal~Moolchandani,
        William~Crossley,
        and~Daniel~DeLaurentis
\thanks{K. Ogunsina, H. Chao, N. Kolencherry, S. Jain, and K. Moolchandani are graduate researchers in the School of Aeronautics and Astronautics, Purdue University.}
\thanks{D. DeLaurentis and W. Crossley are Professors of Aeronautics and Astronautics, Purdue University.}
\thanks{Manuscript received February 2, 2018; revised April 24, 2018.}
}

\maketitle

\begin{abstract}
The International Air Transport Association (IATA) is one of several organizations that have presented goals for future $CO_2$ emissions from commercial aviation with the intent of alleviating the associated environmental impacts. These goals include attaining carbon-neutral growth in the year 2020 and total aviation $CO_2$ emissions in 2050 equal to 50\% of 2005 aviation $CO_2$ emissions. This paper presents the use of a simulation-based approach to predict future $CO_2$ emissions from commercial aviation based upon a set of scenarios developed as part of the Aircraft Technology Modeling and Assessment project within ASCENT, the FAA Center of Excellence for Alternative Jet Fuels and the Environment. Results indicate that, in future scenarios with increasing demand for air travel, it is difficult to reduce $CO_{2}$ emissions in 2050 to levels equal to or below 2005 levels, although neutral $CO_{2}$ growth after 2020 may be possible. 
\end{abstract}


%
\IEEEpeerreviewmaketitle

\input{Introduction}

\input{Scenario_Description}

\input{Modeling_Tools}

\input{Simul_Results}

\input{Conclusion}

\section*{Acknowledgment}
This work was funded by the US Federal Aviation Administration (FAA) Office of Environment and Energy as a part of ASCENT Project 10 under FAA Award Number: 13-C-AJFE-PU. Any opinions, findings, and conclusions or recommendations expressed in this material are those of the authors and do not necessarily reflect the views of the FAA or other ASCENT Sponsors. 

\ifCLASSOPTIONcaptionsoff
  \newpage
\fi



%
\bibliography{ref_hsun,ref_samarth}
\bibliographystyle{ieeetr}

%


\begin{IEEEbiographynophoto}{Kolawole Ogunsina}
is a second-year Ph.D. student in the Center for Integrated Systems in Aerospace (CISA) laboratory at the Purdue University School of Aeronautics and Astronautics, under the distinguished advisement of Dr. DeLaurentis. His research areas are in Aerospace Systems; primarily the exploration of new concepts (such as blockchain technology) for addressing airline disruption management, and appraising the impact of airline operations on future environmental emissions.
\end{IEEEbiographynophoto}
\begin{IEEEbiographynophoto}{Hsun Chao}
joined Purdue University in 2014 as a Master's student. He received his B.S. in Physics and Aeronautics and Astronautics from National Cheng Kung University in Tainan, Taiwan, and his M.S. in Aeronautics and Astronautics from Purdue University. He is currently a Research Assistant in the Center for Integrated Systems in Aerospace (CISA) laboratory at Purdue University, with Dr. DeLaurentis as his advisor. His research interests are in the national air traffic system, airline operations, and distributed systems-of-systems.
\end{IEEEbiographynophoto}
\begin{IEEEbiographynophoto}{Nithin Kolencherry}
is a graduate student in Aeronautics and Astronautics at Purdue University, where he has been working on his Ph.D. since 2015. His research interests are in the areas of multi-fidelity design optimization and aircraft/spacecraft conceptual design. 
\end{IEEEbiographynophoto}
\begin{IEEEbiographynophoto}{Samarth Jain}
is a graduate research assistant, pursuing his Masters in Aeronautics and Astronautics at Purdue University. He received his Bachelors in Mechanical and Automotive Engineering from Delhi Technological University, India in 2016. His research interests are in multidisciplinary design optimization and aircraft conceptual design.
\end{IEEEbiographynophoto}
\begin{IEEEbiographynophoto}{Kushal Moolchandani}
is a graduate research assistant in the School of Aeronautics and Astronautics at Purdue University, working towards his Ph.D. degree under Dr. Daniel DeLaurentis. His research interests include aircraft design and optimization, air transportation systems, systems-of-systems, and autonomous systems development. He got his bachelor's degree from the PEC University of Technology, Chandigarh, India, and his Master's degree at Purdue, both in in aeronautical engineering.  In the area of air transportation systems he worked on the development of the Fleet-Level Environmental Evaluation Tool (FLEET) which also formed part of this Master’s thesis. Currently, he is working on a project to study the effects of the decision-making of autonomous nodes on complex network performance.
\end{IEEEbiographynophoto}
\begin{IEEEbiographynophoto}{William Crossley}
is a Professor of Aeronautics and Astronautics at Purdue University, where he has been a member of the faculty since August 1995. His teaching and research interests are in optimization for systems and system-of-systems problems, with an emphasis on aerospace and aviation applications.
\end{IEEEbiographynophoto}
\begin{IEEEbiographynophoto}{Daniel DeLaurentis}
is a Professor of Aeronautics and Astronautics at Purdue University. He joined the university as a strategic faculty hire in Purdue's Signature Area for System-of-Systems (SoS) research in August 2004. His research is focused upon the development of foundational methods and tools for addressing problems characterized as system-of-systems. His present context for the research is the exploration of Next-generation Air Transportation systems, especially including the presence of revolutionary aerospace vehicles, new business models, and alternate policy constructs.
\end{IEEEbiographynophoto}






\end{document}

%% file: Introduction.tex
\section{Introduction}

\IEEEPARstart{T}{o} encourage the reduction of negative environmental impact from an important engineering system -- in this case, air transportation -- goals that set challenging targets for future environmental impact can be important. This paper assesses differences between predicted carbon emissions from a simulation tool that models numerous aspects of commercial aviation and the carbon emission reduction goals from the International Air Transport Association (IATA). These future $CO_2$ predictions use various economic conditions and technology improvement scenarios; other than a simple carbon pricing model, the scenarios do not incorporate other policies.

\begin{figure}[tbh]
	\centering
    \includegraphics[width=\linewidth]{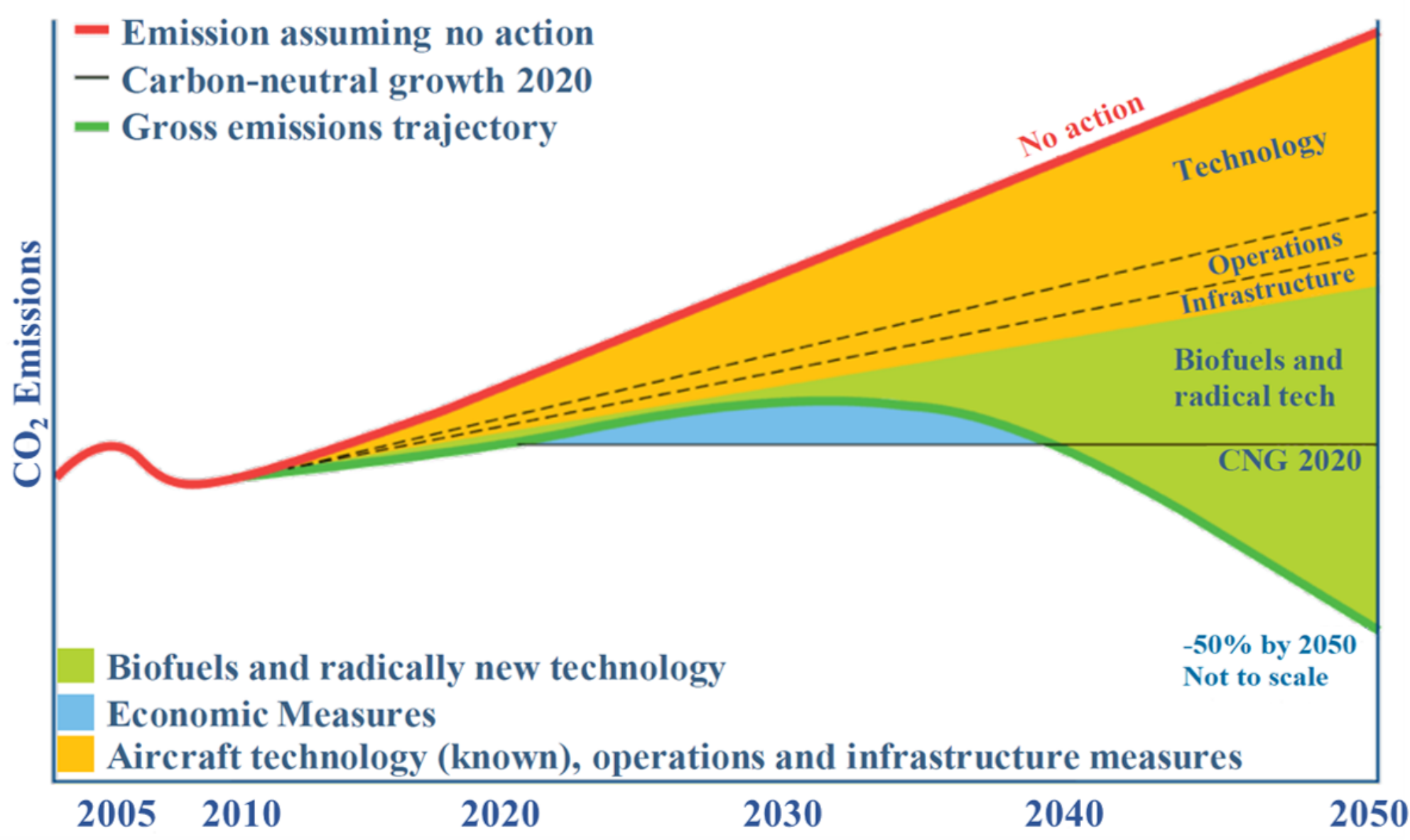}
    \caption{Aviation $CO_2$ emissions, envisioned reductions, and IATA goals\cite{SteeleAEmissions}.}
    \label{f:IATA Goal}
\end{figure}

IATA -- which represents airlines and over 80\% of worldwide aviation activity, and whose formation predates that of International Civil Aviation Organization (ICAO) -- provides technical and other input for airline regulation and is charged by governments to regulate various aspects of international aviation. IATA recently called for a greater urgency in the partnership between airlines and governments to ensure aviation remains on the leading front for industries in sustainably managing the effect of climate change\cite{IATACORSIA}. IATA believes that an airline-government partnership is important for creating feasible and acceptable policies for reducing the environmental impact of commercial aviation and also vital to setting goals to evaluate the efficacy of those policies. The current IATA reduction goals include carbon neutral growth from 2020 and cutting net $CO_2$ emission to half of their 2005 values by 2050\cite{IATACORSIA,SteeleAEmissions}. Fig.~\ref{f:IATA Goal} presents these goals. 

The research method consists of two steps. In the first step, the authors participated with their partners in ASCENT (the Aviation Sustainability Center), the FAA Center of Excellence for Alternative Jet Fuels and Environment, working on the project ``Aircraft Technology Modeling and Assessment'' to define several future commercial aviation scenarios \cite{DelaurentisProjectSummary}. These scenarios emerged from the data obtained from participants of a two-phase workshop. Workshop participants helped to define descriptors that would describe future scenarios of the U.S. commercial aviation industry; these included Gross Domestic Product (GDP) growth rate, population growth rate, energy or fuel prices, and rate of aircraft technology development.  For each of these descriptors, the workshop participants also helped set nominal, minimum and maximum levels.  Each future scenario uses a different combination of these descriptor levels.  The ASCENT Aircraft Technology and Modeling Assessment team selected combinations that sought to span a broad range of possible future outcomes.

In the second step, the authors of this paper performed simulations of these future scenarios for commercial aviation in the U.S. using Fleet-Level Environmental Evaluation Tool (FLEET). FLEET can assess environmental impact of commercial aviation in the U.S., particularly fleet-level $CO_2$ emissions,  with considerations of advanced concept of aircraft, airline operations, and eco-friendly policies for the U.S. commercial aviation industries \cite{Moolchandani2011,Moolchandani2012,Moolchandani2013,Moolchandani2016,Chao2016b,Chao2016,Chao2016AirlineApproach,Chao2017SensitivityAirlines,Moolchandani2017}. The backbone of FLEET is a resource allocation problem that mimics the behavior of airlines assigning their fleet of aircraft on each route in their networks. Using a series of modules that represent various aspects of the commercial aviation system connected by feedback loops, FLEET allows airline ticket fares, passenger demand, airline fleet size and airline fleet composition to evolve over time.

By including routes and operations that reflect most of the ``U.S. touching'' airline operations (i.e., routes that have at least one airport in the U.S.), the results from FLEET provide a prediction of the $CO_2$ emissions trends from U.S. commercial aviation. Comparing the carbon emission trends from the simulation results with the IATA emission reduction goals, revealed that -- for the scenarios studies here that emphasize potential future economic growth and potential future technology development -- the predicted $CO_2$ emissions do not quite reach the levels specified by the IATA goals.

%% file: Scenario_Description.tex
\section{Future Scenario Description}

\begin{figure}[hbt]
	\centering
    \includegraphics[width=0.99\linewidth]{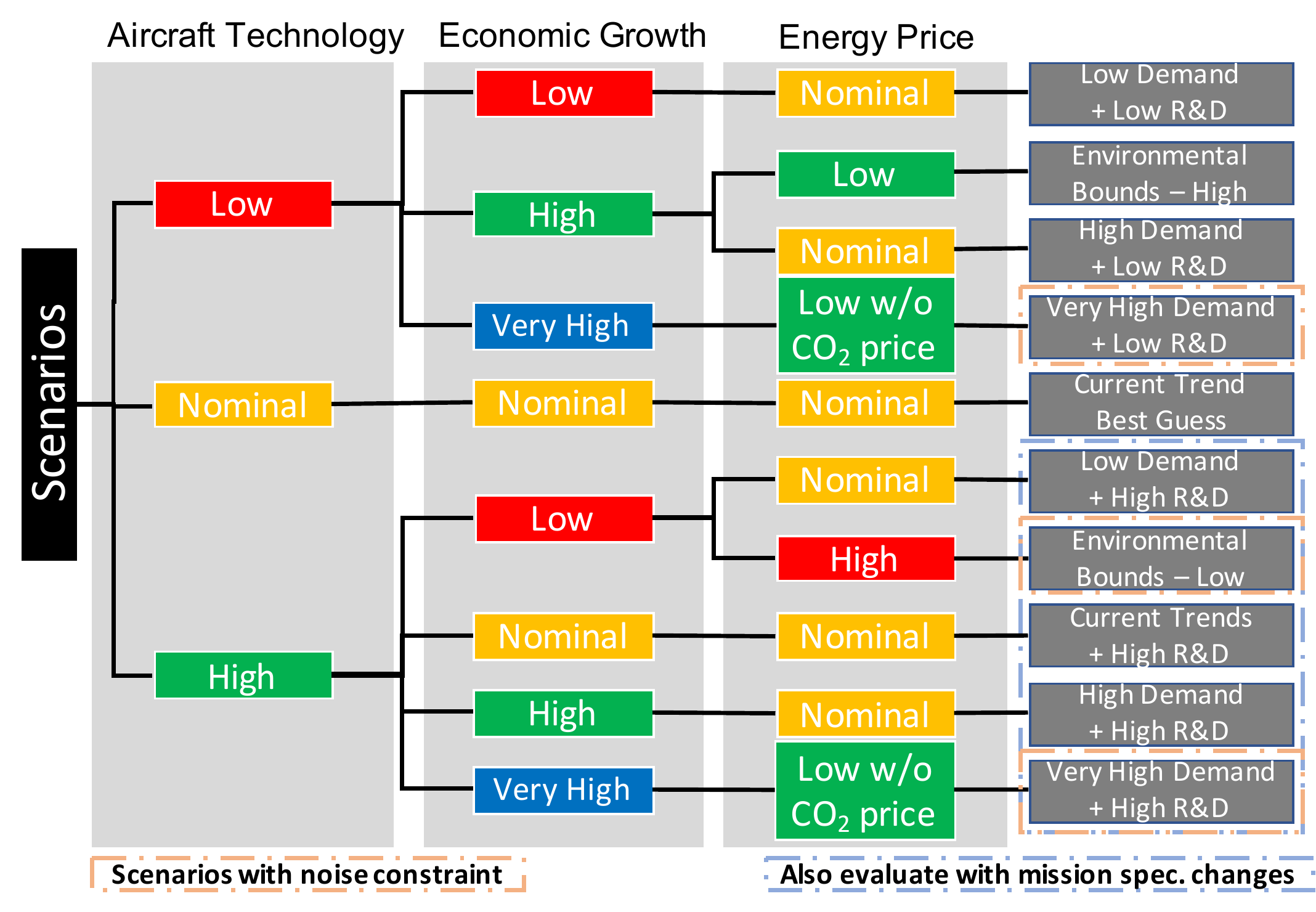}
    \caption{Scenario Tree Overview}
    \label{f:Scenario Tree}
\end{figure}

To generate the potential future commercial aviation scenarios, the ASCENT Aircraft Technology Modeling and Assessment project team conducted two phases of workshops that gathered feedback from several industry, government and academic participants to select scenario descriptors, range of values for those descriptors and the importance of those descriptors. This effort led to developing scenarios by combining different levels of aircraft technology development, economic growth, and energy price, as Fig.~\ref{f:Scenario Tree} presents. Other possible scenario combinations not represented in Fig.~\ref{f:Scenario Tree} were deemed impractical by workshop participants. 

The Aircraft Technology levels describe the predicted performance, particularly fuel consumption, of future aircraft and engine combinations as well as the Entry Into Service (EIS) dates for these future aircraft.  A ``High'' level indicates future aircraft with very good performance becoming available to the airline soon. ``Nominal'' aircraft technology reflects performance levels and EIS dates that follow a consensus of the current aircraft and engine development trajectory. A ``Low'' aircraft technology level indicates future aircraft with poorer performance than those of the ``Nominal'' level; however, these aircraft will still show improvements over today's generation of aircraft.  While there is great interest in ideas like biofuels to reduce life-cycle $CO_2$ emissions and electric or hybrid electric propulsion that might make more carbon-efficient means to provide aircraft thrust and power, the aircraft technology level here does not distinguish these specific technologies.  All of these are reflected in an equivalent level of improved fuel consumption in the future aircraft models.  This means that, where Fig.~\ref{f:IATA Goal} shows potential improvements from ``technology'' and ``biofuels and radical technology'' separately, the simulations here treat these as aggregated technology-driven improvements.

The Economic Growth descriptors illustrate the economic condition around the world, primarily described by Gross Domestic Product (GDP) growth rate. The descriptor influences the amount of passenger demand and distribution of this demand across the airline network. In this study, authors considered GDP growth rates for countries in North America, South America, Europe, Africa, Asia, and Oceania. As shown in Table~\ref{t:GDP Continents}, ``Very High'' and ``High'' levels indicate the same GDP growth rates for the countries in the six continents. The very high setting has no carbon pricing in addition to high GDP growth rates, which will lead to the highest possible demand. A ``Nominal'' level indicates lower GDP growth rate for each continent, while a ``Low'' level indicates the lowest GDP growth rate. 

\begin{table*}[thbp]
  \centering
  \caption{GDP Growth Rate in Continents}
  \label{t:GDP Continents}
  \begin{tabular}{|c|c|c|c|c|c|c|}
    \hline
     & \textbf{\begin{tabular}[c]{@{}c@{}}North \\ America\end{tabular}} & \textbf{\begin{tabular}[c]{@{}c@{}}South \\ America\end{tabular}} & \textbf{Europe} & \textbf{Africa} & \textbf{Asia} & \textbf{Oceania} \\ \hline
    \textbf{Low} & 1.8 & 2.7 & 0.6 & 1.8 & 3.3 & 1.8 \\ \hline
    \textbf{Nominal} & 2.8 & 4.2 & 2.4 & 2.8 & 4.3 & 2.8 \\ \hline
    \textbf{High/Very High} & 4.0 & 5.3 & 4.2 & 4.0 & 5.9 & 4.0 \\ \hline
  \end{tabular}
\end{table*}

Lastly, the Energy Price descriptors provide different trends for fuel price and any potential carbon emission price trends. The energy price descriptors influence the cost of operating the airline fleet, which cascades through ticket price to a price-elasticity response in demand. The values for the energy price descriptor shown in Table~\ref{t:Scenario_Settings} represent oil prices in 2050. Additionally, the cost of $CO_2$ emission linearly increase from zero $CO_2$ price in 2020 to 2050 to reach the cost of $CO_2$ emission listed in Table~\ref{t:Scenario_Settings} for each scenario.

The Mission Specification Change (MSC) descriptor presents an additional means of mitigating commercial aviation carbon emissions. In Fig.~\ref{f:Scenario Tree}, a blue dashed-dotted line surrounds all of the scenarios that use a high technology level as part of their definition.  For these, a second version of these scenarios considered the potential impact of reducing the cruise speed of the new aircraft acquired by the airline. Designing a new aircraft for a slower cruise speed than today's current aircraft could decrease the carbon emissions per passenger on the new aircraft.  For the aircraft considered in this study, the cruise speed reduction resulted in per flight fuel consumption reductions from 5\% to 15\% below an aircraft designed to fly at speeds consistent with today's transport aircraft.

While the main focus of this effort considered future $CO_2$ emissions, noise around airports is another environmental impact of aviation.  The ``very high demand'' scenarios would lead to a high number of flights to meet the demand.  The ``environmental bounds -- low'' scenario provided a lower bound for predicted $CO_2$ emission; in other words, this scenario should lead to the lowest possible $CO_2$ emissions. For the ``very high demand'' scenarios and the ``environmental bounds -- low'' scenario, the allocation problem in FLEET included constraints to keep the noise level around an airport below a prescribed level; this would often restrict the number of operations at airports with this constraint active. 

The authors decided not to pursue a full factorial-like design of experiments across the varying levels of scenario descriptors. The scenario simulations are moderately expensive computationally, so this effort focused upon a set of available combinations that would bound the potential $CO_2$ emission output. For example, a ``Low'' aircraft technology scenario with a ``Nominal'' economic growth level would likely produce $CO_2$ levels that lie between the ``Low'' aircraft technology with ``Low'' and ``High'' economic growth levels.

\begin{table*}[bhtp]
  \centering
  \caption{Parameter Setting for Simulation Scenarios}
  \label{t:Scenario_Settings}
  \begin{tabular}{|l|c|c|c|c|}
    \hline
    \multirow{2}{*}{} & \multirow{2}{*}{\textbf{\begin{tabular}[c]{@{}c@{}}Aircraft \\ Technology\end{tabular}}} & \textbf{Economic Growth} & \multicolumn{2}{c|}{\textbf{Energy Price}} \\ \cline{3-5} 
     &  & \textbf{\begin{tabular}[c]{@{}c@{}}North America\\ GDP Growth\\ (\%/year)\end{tabular}} & \textbf{\begin{tabular}[c]{@{}c@{}}Energy Price\\ (\$/bbl)\end{tabular}} & \textbf{\begin{tabular}[c]{@{}c@{}}Cost of CO2\\  Emissions (\$/MT)\end{tabular}} \\ \hline
    \textbf{Current Trends Best Guess} & Nominal & 2.8 & 77 & 21 \\ \hline
    \textbf{Current Trends + High R\&D (MSC)} & High & 2.8 & 77 & 21 \\ \hline
    \textbf{Environmental Bounds -- Low (MSC)} & High & 1.8 & 181 & 85 \\ \hline
    \textbf{Environmental Bounds -- High} & Low & 4.0 & 41 & 0 \\ \hline
    \textbf{High Demand + Low R\&D} & Low & 4.0 & 77 & 21 \\ \hline
    \textbf{High Demand + High R\&D (MSC)} & High & 4.0 & 77 & 21 \\ \hline
    \textbf{Low Demand + Low R\&D} & Low & 1.8 & 77 & 21 \\ \hline
    \textbf{Low Demand + High R\&D (MSC)} & High & 1.8 & 77 & 21 \\ \hline
    \textbf{Very High Demand + Low R\&D} & Low & 4.0 & 41 & 0 \\ \hline
    \textbf{Very High Demand + High R\&D (MSC)} & High & 4.0 & 41 & 0 \\ \hline  \end{tabular}
\end{table*}

%% file: Modeling_Tools.tex
\section{Modeling Tool -- FLEET}

FLEET is a computational simulation tool that predicts how fleet-level environmental impacts of aviation -- in the form of $CO_2$ emissions and airport noise -- evolve over time. FLEET follows a system dynamics-inspired approach to connect computational modules that mimic the economics of airline operations, models airlines decisions regarding retirement of old aircraft and acquisition of new aircraft, and represents passenger demand growth in response to economic conditions. Central to FLEET is an allocation problem that represents how an airline would use its aircraft to meet passenger demand to maximize profit. This set of interconnected computational modules enables the tool to assess the impact of future aircraft concepts and technologies on fleet-wide environmental metrics, while reflecting resulting relationships between emissions, passenger demand/market demand, ticket prices, and airline fleet composition from 2005 to 2050. FLEET is capable of providing a prediction of how variation in external factors such as economic conditions, policy implementation, and technology availability would affect future commercial aviation environmental impacts. Currently, FLEET focuses only on passenger demand and airline operations on U.S. domestic routes and international routes that have either their origin or destination in the United States\cite{Isaac09,zhao09,zhao10,mane10,Moolchandani2011,Moolchandani2012,moolchandani12_2,Moolchandani2013}. The FLEET simulations presented in this paper use 2005 as the starting year for all simulations; this is also the baseline year used in several $CO_2$ emissions goals, like those of IATA. For the years 2005 to 2015, FLEET also uses reported values of historical passengers carried from the Bureau of Transportation Statistics (BTS) as the demand. After 2015, FLEET uses the economic and price factors to predict future passenger demand.

In FLEET, four different technology age categories and six different size classes represent the aircraft available to the airline. The four technology age categories are:
\begin{enumerate}
  \item Representative-in-class aircraft are the most flown aircraft in 2005 (base year for FLEET)
  \item Best-in-class aircraft are the ones with most recent entry-in-service dates in 2005, but were in the airline fleet
  \item New-in-class aircraft are either aircraft currently under development that will enter service in the near future or concept aircraft that incorporate technology improvements expected in the near future
  \item Future-in-class aircraft are the generation of aircraft that will enter service after the new-in-class aircraft and will include additional improvements in technology and associated performance.
\end{enumerate}

The aircraft within each technology age category further subdivide into as many as six classes, based upon seat capacity.  These classes represent the mix of aircraft sizes in the airline fleet.  For the representative- and best-in-class aircraft, the six FLEET aircraft classes are: 1) Small Regional Jet up to 50 seats (SRJ), 2) Regional Jet, 3) Small Single Aisle, 4) Large Single Aisle, 5) Small Twin Aisle, and 6) Large Twin Aisle. FLEET uses five new- and future-in-class aircraft classes numbered from 2 to 6 and leaves class 1 empty, recognizing that new orders for future 50-seat regional jets have diminished to zero.  The FLEET new- and future-in-class divisions are: 2) Regional Jet (RJ), 3) Single Aisle (SSA-LTA), 4) Small Twin Aisle (STA). 5) Large Twin Aisle (LTA), and 6) Very Large Aircraft (VLA).

Table \ref{aircraft_type} shows the various aircraft used in the FLEET simulations.  The representative- and best-in-class aircraft are marked with their respective aircraft labels, and the new- and future-in-class aircraft are marked with their corresponding EIS dates used in the study. In Table \ref{aircraft_type} , the aircraft labeled with “Gen1 DD” and “Gen2 DD” are the Generation 1 and Generation 2 aircraft respectively, both with an advanced ``Direct Drive'' turbofan engine. These include aircraft that belong to the following classes -- regional jet (RJ), single aisle (SSA-LSA), small twin aisle (STA), large twin aisle (LTA), and very large aircraft (VLA). The New-in-Class and Best-in-Class aircraft models vary on the basis of the amount and speed of technology incorporated into aircraft in each of the scenarios. Details of these aircraft are available in Ref. \cite{DelaurentisProjectSummary}.

\begin{table}
\caption{Aircraft Types in Study with [Label] and (EIS)}\label{aircraft_type}
\begin{tabular}{ | p{0.9cm} | p{1.6cm} | p{1.5cm} | p{1.3cm} | p{1.3cm} | }
\hline
{} & Representative-in-class & Best-in-class & New-in-class & Future-in-class \\
\hline
Class 1 & Canadair RJ200/RJ440 [SRJ] & Embraer ERJ145 [SRJ] &  & \\
\hline
Class 2 & Canadair RJ700 [RJ] & Canadair RJ900 [RJ] & Gen1 DD RJ (2020) & Gen2 DD RJ (2030) \\
\hline
Class 3 & Boeing 737-300 [SSA] & Boeing 737-700 [SSA] & Gen1 DD SSA-LSA (2017) & Gen2 DD SSA-LSA (2035) \\
\hline
Class 4 & Boeing 757-200 [LTA] & Boeing 737-800 [LTA] & Gen1 DD STA (2025) & Gen2 DD STA (2040) \\
\hline
Class 5 & Boeing 767-300ER [STA] & Airbus A330-200 [STA] & Gen1 DD LTA (2020) & Gen2 DD LTA (2030) \\
\hline
Class 6 & Boeing 747-400 [LTA] & Boeing 777-200LR [LTA] & Gen1 DD VLA (2025) & Gen2 DD VLA (2040) \\
\hline
\end{tabular}
\end{table}

The aircraft allocation problem central to FLEET seeks to maximize airline profit while meeting demand and satisfying operational constraints. The airline profit is simply the difference between revenue and cost. The revenue is a function of the ticket price and number of passengers on every route for each aircraft type. The ticket price depends on the aircraft type and route on which the passenger flies. The cost is a function of the direct operating cost and the number of trips of each aircraft type for every route. Thus, the total airline profit is the sum of profit from every route and for each of the aircraft type. 

The allocation problem has four types of operational constraints. First, the number of passengers transported by the airline on a route should not exceed the passenger demand for that route. Second, the airline should meet at least 20\% of the passenger demand on each route; without explicit competition in the model, this lower limit prevents the FLEET airline from ``dropping'' a route.  For most FLEET solutions, the airline serves 95\% or more of the demand. Third, the number of each type of aircraft operated by the airline is limited to the number of aircraft owned by the airline; this is enforced by ensuring the number of hours needed for each type of aircraft -- flying the route, turnaround at airports, and down time for maintenance -- in three days is less than 72 hours multiplied by the number of that type of aircraft owned by the airline.  The three day period allows for round-trip, trans-Pacific flights. Fourth, a constraint ensures that the number of passengers carried on a given type of aircraft does not exceed the number of seats in that aircraft type. 

The aircraft allocation problem is solved using the General Algebraic Modeling System (GAMS) software package \cite{gamssoftware} and the IBM CPLEX solver\cite{InternationalBusinessMachinesCorp.2014}. Using the number of trips of each aircraft type allocated to each route in the solution, FLEET determines the fuel burn (which relates directly to $CO_{2}$) for each aircraft on every route in the airline network. A detailed description of the allocation problem formulation and the initial fleet, airport, and network setup can be found in Ref. \cite{Moolchandani2017}.

FLEET uses a nonlinear relationship to model the demand growth rate in different continents (see Ref. \cite{airbusGMF14}), which is based on the historical data of trips/capita vs. GDP/capita. The data implies that if all the continents had the same GDP growth rate, the continents with higher GDP/capita would have a lower trips/capita growth rate. 

The model employs the GDP and population data of each continent in 2005 from World Bank \cite{population} as initial settings. By using the GDP growth rate and population growth rate historical data and predictions, the model tracks the demand for each continent from 2005 to 2050 simulation year.

FLEET uses total noise area as a fleet-level metric to model noise around airports. The total noise area metric is the sum of the predicted area inside the 65 dB day-night average sound level contour at all the noise-limited airports in the FLEET network, and it serves as a single metric to describe the broad fleet impact. A fleet allocation with a larger total noise area would indicate more fleet-level noise. This metric is used as a hard constraint for the noise-limited scenarios, ensuring that the airline does not produce more total noise area beyond the noise area limit. This limits the number of flight operations across all of the noise-limited airports. The noise-limited airports do not include international airports because the airline model does not attempt to represent a significant
number of operations at those airports; the current airline model more nearly represents all operations at U.S. airports \cite{moolchandani17}.

%% file: Simul_Results.tex
\section{Simulation Results}

This section describes the $CO_{2}$ emissions -- obtained from the FLEET simulation from 2005 to 2050 -- for all the different scenarios as a function of variations in economic growth and aircraft technology level. 

\subsection{Variation in Economic Growth}
\begin{figure*}[htbp]
	\centering
    \subfloat[Variation of $CO_{2}$ Emissions across all Scenarios as a function of Economic Growth (Note that only the ``Very High'' scenarios include the noise area constraint) ]{\includegraphics[width=0.99\linewidth]{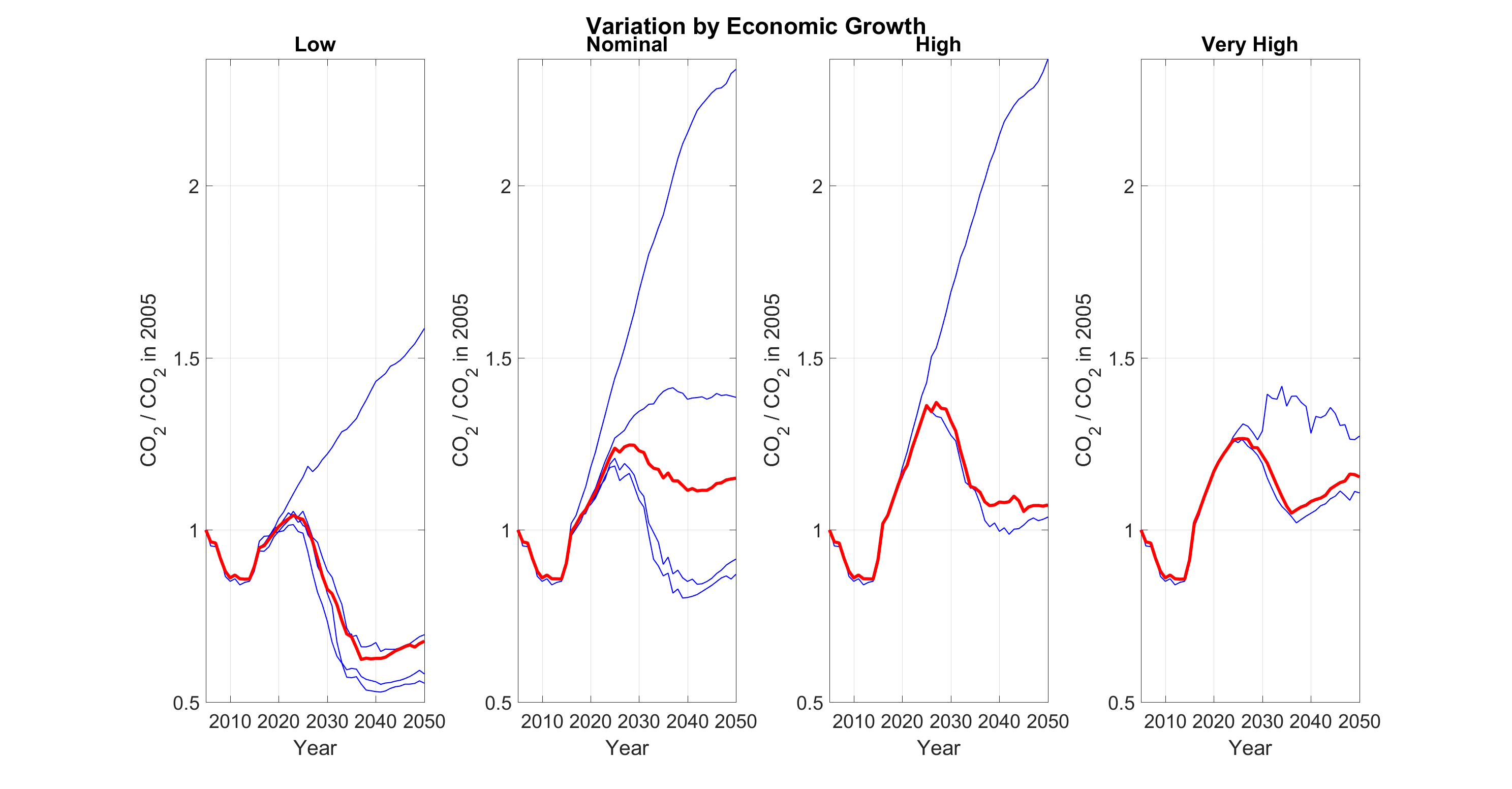}\label{f:Variation by Growth}}\\
    \subfloat[Variation of $CO_{2}$ Emissions across all Scenarios as a function of Aircraft Technology Level]{\includegraphics[width=0.99\linewidth]{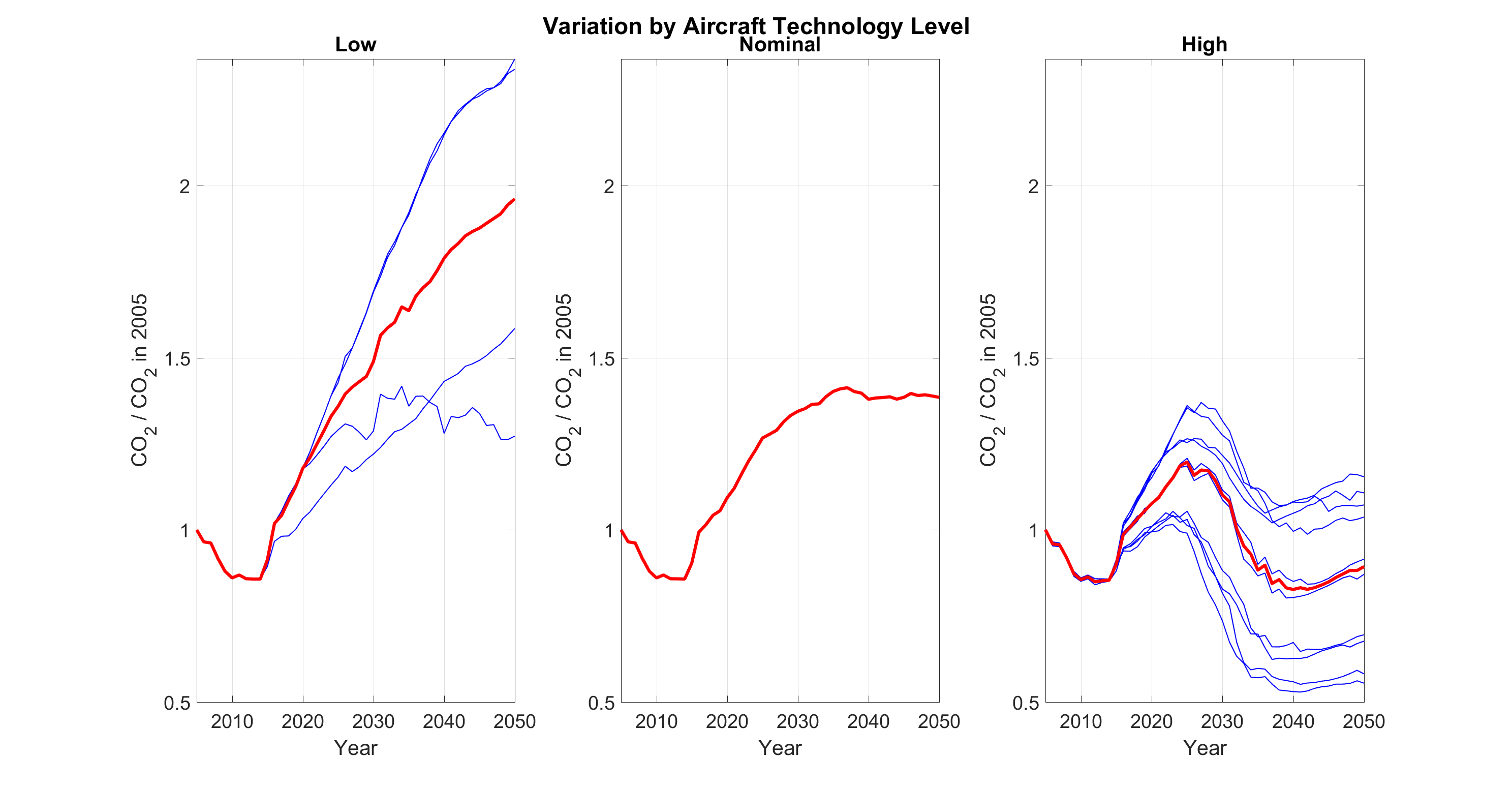}\label{f:Variation by Tech}}
   	\caption{Simulation Results}\label{f:Results}
\end{figure*}

Fig.\ref{f:Variation by Growth} shows the normalized $CO_{2}$ emissions from the FLEET airline during the 46-year simulation period for all scenarios with regards to their respective variations in economic growth. The figure classifies the scenarios into ``Low'', ``Nominal'', ``High'', and ``Very High'' economic growth conditions. ``Low'' is representative of potential future scenarios with lower economic growth than current trends in economic growth; there are five scenarios with low economic growth level. ``Nominal'' includes three future scenarios where current economic growth trends are upheld. While ``High'' and ``Very High'' classes of scenarios have similar economic growth levels, the ``Very High'' class of scenarios are subject to airport noise constraints and have significantly lower energy prices than the ``High'' class of scenarios. The ``High'' and ``Very High'' classes of scenarios include four and three different scenarios, respectively. The total $CO_{2}$ emissions from the airline in 2005 is used to normalize the $CO_{2}$ emissions throughout the simulation period in all four classes of scenarios. The thick red line represents the median (i.e. middle point) values of normalized $CO_{2}$ emissions throughout the simulation period for each class of scenarios. The blue lines surrounding the median line represent different specific scenarios within each classification of economic growth, such that the range of variation of $CO_{2}$ emissions is represented by the gap between lines with the highest and lowest values of $CO_{2}$ emissions throughout the simulation period. 

As evidenced in Fig.\ref{f:Variation by Growth}, the median normalized $CO_{2}$  for the class of scenarios with low economic growth decreased to about 70\% of the  value in 2005. This suggests that half of the scenarios with low economic growth resulted in more than 0.7 times of the 2005 $CO_{2}$ emission value by 2050. The median normalized $CO_{2}$ emission by 2050 for the classes of scenarios with moderately and significantly high economic growth increased to 1.05 times and 1.2 times, respectively, of $CO_{2}$ emission value in 2005. The class of scenarios with nominal and high economic growth resulted in the largest range of variation of $CO_{2}$ emission (difference of about 150\% of the 2005 $CO_{2}$ emission between the upper and lower bounds by 2050) for all classes of scenarios. This reveals that the evolution of $CO_{2}$ emissions from early 2030s to 2050 are very sensitive to the availability of next generation aircraft and energy prices in scenarios where economic growth is high.

While the class of scenarios with high economic growth have a large variation in $CO_{2}$ emissions between the mid 2020s and 2050, the variation in $CO_{2}$ emissions during this period is asymmetric such that the corresponding median $CO_{2}$ emissions are very close to the lower bound. Fig.\ref{f:Variation by Growth} shows that half of the scenarios with high economic growth yielded similar $CO_{2}$ emissions values between the mid 2020s and 2050 (105\% of 2005 $CO_{2}$ emissions by 2050), and the remaining half of these scenarios yielded $CO_{2}$ emissions that varied significantly (up to 250\% of 2005 $CO_{2}$ emissions by 2050) from other scenarios with high economic growth. Although the class of scenarios with very high economic growth yielded a similar trend (median $CO_{2}$ emission values between the mid 2020s and 2050 are close to the lower bound) as the class of scenarios with high economic growth, this class of scenarios yielded a significantly smaller range of variation when compared to the class of scenarios with high economic growth. 

The considerable reduction in $CO_{2}$ emissions in the ``Very High'' class of scenarios is due to the noise limits imposed on the airline. The noise limitation significantly reduced the number of aircraft operations, which also decreased $CO_{2}$ emissions. The $CO_{2}$ emission starts to increase again after 2035 because the airline receives quieter aircraft with newer technology, thereby increasing its aircraft operations at noise-limited airports. Fig.\ref{f:Variation by Growth} also reveals that low economic growth can produce carbon emission levels below the corresponding 2005 value after 2030, while nominal economic growth can yield carbon emission levels below the 2005 value or carbon neutral growth after 2031. However, none of the scenarios yielded 50\% of the 2005 $CO_2$ emission level by 2050 to achieve IATA $CO_2$ emission reduction goals.

\subsection{Variation in Aircraft Technology Level}

Fig.\ref{f:Variation by Tech} shows the normalized $CO_{2}$ emissions from the airline during the simulation period for all scenarios classified based on variations in aircraft technology level. The scenarios are classified into ``Low'', ``Nominal'', and ``High'' respectively. ``Nominal'' represents the single scenario where only current trends in next generation aircraft are available to the airline (Current Trends Best Guess). ``Low'' and ``High'' represent classes of scenarios where the performance (fuel consumption, non fuel-related operating costs, etc.) of next generation aircraft available to the airline during the simulation are lower and higher, respectively, than the performance of corresponding next generation aircraft in the ``Nominal'' scenario. The ``Low'' and ``High'' classes of scenarios are represented by four and ten different scenarios respectively. Similar to the results in Fig.\ref{f:Variation by Growth}, the total $CO_{2}$ emissions from the airline in 2005 is used to normalize the $CO_{2}$ emissions throughout the simulation period in all four scenario classes. The thick red line represents the median values of normalized $CO_{2}$ emissions throughout the simulation period for each class of scenarios. The blue lines surrounding the median line represent different scenarios for each classification of economic growth, such that the range of variation of $CO_{2}$ emissions is represented by the gap between lines with the highest and lowest values of $CO_{2}$ emissions throughout the simulation period.  
	
As shown in Fig.\ref{f:Variation by Tech}, the class of scenarios with low aircraft technology level has the highest median normalized $CO_{2}$ emission by 2050, reaching about 200\% of the corresponding 2005 value. This affirms that half of the scenarios with low aircraft technology resulted in less than 200\% of the 2005 $CO_{2}$ emission value by 2050. The increasingly large range of $CO_{2}$ emissions from 2020 to 2050 for the class of scenarios with ``Low'' aircraft technology level suggests that $CO_{2}$ emissions by 2050 are very sensitive to economic growth and energy prices when research and development of next generation aircraft are low. The Current Trends Best Guess scenario (``Nominal'') resulted in a 40\% increase above the 2005 $CO_{2}$ emission value in 2050. The nearly constant $CO_{2}$ emission observed between 2038 and 2050 in the ``Nominal'' scenario reflects the attenuating effect of the available next generation aircraft, in presence of increasing economic growth and energy prices, on $CO_{2}$ emissions. The median normalized $CO_{2}$ emissions in the class of scenarios with high aircraft technology decreased to 85\% of the 2005 value by 2050. The decreasing trend observed in $CO_{2}$ emissions between the mid 2020s and mid 2030s in the ``High'' class of scenarios is due to the introduction of very fuel-efficient next generation aircraft made possible by high level aircraft research and development. However, as economic growth continues to increase in the class of scenarios with high aircraft technology, $CO_{2}$ emissions slowly start to increase between 2040 and 2050 as demand starts to overwhelm the technology improvements. Fig.\ref{f:Variation by Tech} also reveals that high research and development of next generation aircraft is necessary to achieve carbon emission levels below the corresponding 2005 value after 2030, while carbon neutral growth can be achieved from the mid 2020s with the introduction of high level aircraft technology.

%% file: Conclusion.tex
\section{Future Work and Conclusions}
The preceding results and discussion demonstrate the ability of FLEET to make model-based predictions of how commercial aviation $CO_{2}$ emissions might evolve from 2005 to 2050 for a variety of scenarios. As with any predictive model, the results are only as good as the underlying models and assumptions in the scenarios. A few opportunities for future work exist here.  

The scenarios studied here utilized four technology ages of aircraft, with the future-in-class age having first EIS in 2030. Given the typical 10 to 15 year development cycle for new aircraft, a ``future-future-in-class'' age of aircraft might have EIS dates around 2045; with further improvements in fuel consumption, perhaps such aircraft might propagate into the airline fleet enough to mitigate some of the rising $CO_{2}$ trends near the end of the scenario simulations shown in Fig.~\ref{f:Results}.  
Other than the indirect effect of the noise area constraints on airport capacity, the results presented in this work assume that the FLEET airline operations are not limited by airport capacity; i.e., the airline is capable of performing as many operations as desired. At the passenger demand growth levels associated with these scenarios, an effort to reflect airport capacity constraints would reflect an additional concern.

The presentation of notional $CO_{2}$ emissions associated with the IATA goals in Fig.~\ref{f:IATA Goal} includes a portion of the $CO_{2}$ reduction associated with ``economic measures''. The FLEET simulations include only a simplistic carbon pricing scheme and no other policies or economic measures.

Even considering these limitations, the predicted U.S. commercial $CO_{2}$ emissions show trends that allow interesting comparisons to the IATA goals. The median $CO_{2}$ emissions across all of the scenarios viewed in the categories of economic growth levels, shown in Fig.\ref{f:Variation by Growth}, display a similarity in overall shape. There is an increase in $CO_{2}$ emissions starting in the first year of using predictions of demand (post 2015) that reach a peak in the 2020s, followed by a decline in $CO_{2}$ whose slope differs based upon the economic level, then finally an upward turn at a later date and with a slope that also varies with the economic level. Looking at these median values, the $CO_{2}$ emission predictions for the array of scenarios here suggest that may be close to, but do not meet, the IATA goals.  

The nearer-term goal of carbon neutral growth by 2020 does not appear in the scenarios. Considering the peak in median emission trends across the economic growth levels as a point where carbon neutral growth is reached (i.e., passenger air travel demand continues to increase, while fleet-level $CO_{2}$ stays constant or decreases), there is possibility for this carbon neutral point between about 2025 and 2030. The scenarios here also fall short of the longer-term IATA goal of 2050 $CO_{2}$ emissions at a level that is 50\% of the 2050 $CO_{2}$ emissions. Only the ``Low'' and ``Nominal'' economic level scenarios have results where the 2050 fleet-level $CO_{2}$ is below the 2005 level. The ``Low'' economic growth condition is undesirable for many reasons, and, in the ``Nominal'' condition, the median 2050 $CO_{2}$ level exceeds that of the 2005 $CO_{2}$. Also apparent in considering the individual scenarios is that there is a reasonable chance that 2050 fleet-level $CO_{2}$ will greatly exceed the 2005 level.

The similar shape of the median $CO_{2}$ emissions across the economic growth levels also echoes the shape of the ``gross emissions trajectory'' presented by IATA, with the FLEET results deviating in the upturn in median $CO_{2}$ emissions that occurs between 2035 and 2045 depending upon the economic growth level. Considering the FLEET results in this context suggest that something like the IATA indicated ``economic measures'' would need to supplement the potential gross emissions trajectories.

When viewing the same scenarios categorized by aircraft technology level shown in Fig.\ref{f:Variation by Tech}, the FLEET results point out the significance of the technology level in approaching the IATA goals. In these results, the median $CO_{2}$ emissions associated with ``High'' aircraft technology show a carbon neutral growth peak in 2025 and a 2050 level about 75\% of the 2005 value.  Again, only the ``High'' aircraft technology scenarios have results with 2050 fleet-level $CO_{2}$ emissions below the 2005 level.  

Clearly, meeting the IATA $CO_{2}$ emissions goals will be challenging and much additional work must be done. In the scenarios presented here, the IATA goals are not attained, but the simulation results do show overall shape similar to the gross emissions trajectory suggested by IATA, signifying the general ideas of the IATA goals are consistent with model-based predictions of FLEET. Of the factors considered here, the aircraft technology level has the most significance in approaching the future aviation $CO_{2}$ emissions goals.